\documentclass[pra,showpacs,twocolumn,showkeys]{revtex4-1}
\usepackage{graphics}
\usepackage{amsmath}
\usepackage{color}
\usepackage{dsfont}
\usepackage[left=0.5in, right=0.5in, top=1in, bottom=1in]{geometry}

\usepackage{float}
\usepackage{graphicx}

\begin{document}
\newcommand{\bibs}{C:/Users/seanm_000/Dropbox/References/BibFile}
\title{Fundamental limits on the electro-optic device figure of merit}
\author{Sean Mossman, Rick Lytel, and Mark G. Kuzyk}
\affiliation{Department of Physics and Astronomy, Washington State University, Pullman, Washington  99164-2814 \\ \today}

\begin{abstract}
Device figures of merit are commonly employed to assess bulk material properties for a particular device class, yet these properties ultimately originate in the linear and nonlinear susceptibilities of the material which are not independent of each other.  In this work, we calculate the electro-optic device figure of merit based on the half-wave voltage and linear loss, which is important for phase modulators and serves as the simplest example of the approach. This figure of merit is then related back to the microscopic properties in the context of a dye-doped polymer, and its fundamental limits are obtained to provide a target.  Surprisingly, the largest figure of merit is not always associated with a large nonlinear-optical response, the quantity that is most often the focus of optimization.  An important lesson to materials design is that the figure of merit alone should be optimized.  The best device materials can have low nonlinearity provided that the loss is low; or, near resonance high loss may be desirable because it is accompanied by resonantly-enhanced, ultra-large nonlinear response so device lengths are short.  Our work shows which frequency range of operation is most promising for optimizing the material figure of merit for electro-optic devices.
\end{abstract}

\maketitle
\section{Introduction}

The fundamental limits on the nonlinear-optical response have contributed to the understanding of fundamental light-matter interactions as well as guided material design intent on higher nonlinear-optical responses.\cite{kuzyk13.01}  In molecular materials such as organic crystals\cite{Zyss82.01} and dye-doped polymers\cite{singe86.01}, the bulk material's properties are determined from an ensemble average over molecular properties allowing a clear connection between quantum and bulk properties. For example,  the hyperpolarizability tensor $\beta$, originating from quantum effects in a molecule, is related to the bulk second-order susceptibility according to\cite{singe87.01}
\begin{equation}\label{eq:micro-macro}
\chi^{(2)} = N \left< \beta^* \right>,
\end{equation}
where $\beta^*$ is the dressed hyperpolarizability, which takes into account the total local field within the material, brackets denote an ensemble average over the active molecules, and $N$ is the molecular number density.

Electro-optic devices generally require materials with large responses to an applied voltage while simultaneously maintaining low loss.  Typical devices operate off-resonance to minimize absorption, making the useful length of the device longer, thus lowering the required switching voltage\cite{liu15.01, lytel92.01}.  To maximize response requires making use of resonant enhancement, which in turn increases the loss but will shrink the device. It is essential to understand the scaling of these two competing effects in designing next-generation electro-optic devices.

The electro-optic figure of merit we take for this work is composed of the two competing quantities of interest for device design: the half-wave voltage and the signal loss. The electro-optic figure of merit is inversely proportional to the product of the half-wave voltage $V_\pi$ and the total linear loss $\Lambda$ over the length of the device\cite{enami16.01, gill12.01}. In a molecular material, these bulk properties are proportional to the molecular responses, so the bulk device figure of merit is ultimately limited by the quantum properties of the constituent molecules.

The phenomena of interest originate in the constitutive equations between the applied electric fields, $\vec{\mathcal{E}}$ and the polarization, $\vec{P}$.  To second-order in the electric field, the polarization in the frequency domain is given by,
\begin{equation}\label{eq:polarization}
P_i^\omega = \chi_{ij}^{(1)} (- \omega; \omega) \mathcal{E}_j^{\omega}  + \chi_{ijk}^{(2)} (- \omega; \omega, 0) \mathcal{E}_j^{\omega} \mathcal{E}_k^0 ,
\end{equation}
where $\mathcal{E}_j^{\omega}$ is the $j^{th}$ Cartesion component of the optical field vector, which is assumed monochromatic and of frequency $\omega$, $\mathcal{E}_k^0$ is the $k^{th}$ component of the applied static field, which modulates the phase of the the optical fields, and the frequency dependence of the susceptibilities by convention represents the outgoing field with a negative frequency and the incident fields are to the right of the semicolon.  Summation convention is assumed, so repeated indices are summed over the three Cartesian components.

The tensor nature of the susceptibilities\cite{jerph78.01} embodies the fact that the polarization need not be along the applied electric field, but can in general be induced at an arbitrary angle to the applied electric field.  Since we are interested in the largest component of the nonlinear response, which will generally correspond to the configuration where all the applied electric fields are aligned with one of the principal axes of the susceptibilities, we will assume this to be the case and replace Eq. \ref{eq:polarization} with the scaler form,
\begin{equation}\label{eq:polarization-scaler}
P^\omega = \chi^{(1)} (- \omega; \omega) \mathcal{E}^{\omega}  + \chi^{(2)} (- \omega; \omega, 0) \mathcal{E}^{\omega} \mathcal{E}^0 .
\end{equation}
Eq. \ref{eq:polarization-scaler} assumes that the principle axis chosen is the one that gives the largest nonlinear response.

The assumptions leading to Eq. \ref{eq:polarization-scaler} may be untrue.  For example, it is possible to design a material in which the principle axes of $\chi^{(1)}$ and $\chi^{(2)}$ are not aligned.  We argue that in such cases, the figure of merit will generally not be as large.  For example, consider the case where the light is polarized along the axis with the largest $\chi^{(2)}$. Since the light will then not be along a principle axis of $\chi^{(1)}$, the light's polarization will rotate, so it will only periodically align with the principal axis of largest $\chi^{(2)}$, thus not taking full advantage of the nonlinearity.  A beam polarized along a principle axes of $\chi^{(1)}$ will maintain linear polarization, but its axis will not align with the favorable axis of $\chi^{(2)}$, thus not taking advantage of the large nonlinearity.

One might imagine that clever trickery might be able to take advantage of a material whose principal axes do not align by making a compromise between the lower effective nonlinearity and lower loss that more than compensates to make the figure of merit better.  While these are worthy approaches for squeezing out as much functionality as possible from a material, the most significant gains will most likely be made by designing materials with ideal dispersion characteristics.

Section \ref{sec:EO-Effect} introduces the electro-optic effect and the figure of merit as a function of macroscopic qualities of a material. Section \ref{sec:micro-macro} reviews the connection between the bulk properties of a material and the microscopic properties of the constituent molecules, including orientational order and local field corrections. Section \ref{sec:limits} determines limits on the electro-optic figure of merit by expressing the molecular susceptibilities under the three-level model in terms of scale-invariant molecular transition and energy properties. Finally, Section \ref{sec:fomdispersion} describes the character of the limits of the figure of merit for a variety of operating frequencies in terms of what molecular properties could produce excellent devices given typical device scales.

\section{The electro-optic effect}\label{sec:EO-Effect}

Using Eq. \ref{eq:polarization-scaler}, the electric displacement $D$ is given by
\begin{equation}\label{eq:displacement}
D^\omega = \mathcal{E}^\omega + 4 \pi \left( \chi^{(1)} (- \omega; \omega) \mathcal{E}^{\omega}  + \chi^{(2)} (- \omega; \omega, 0) \mathcal{E}^{\omega} \mathcal{E}^0 \right).
\end{equation}
Using the constitutive relation $D^\omega = \epsilon(\omega) \mathcal{E}^\omega$, we get
\begin{eqnarray}\label{eq:DielectricFunction}
\epsilon(\omega) &=& \left(1 + 4 \pi \chi^{(1)} (- \omega; \omega) \right)  +  4 \pi \chi^{(2)} (- \omega; \omega, 0) \mathcal{E}^0 \nonumber \\
&=& \epsilon^{(0)} (\omega)+ 4 \pi \chi^{(2)} (- \omega; \omega, 0) \mathcal{E}^0 ,
\end{eqnarray}
where $\epsilon^{(0)} (\omega) = 1 + 4 \pi \chi^{(1)} (- \omega; \omega)$ is the linear dielectric function, i.e. when nonlinear effects are absent ($\chi^{(2)} = 0$) or the applied static field vanishes.  Note that since the nonlinear contribution is small by design,
\begin{equation}\label{eq:SmallNonlinearity}
\epsilon^{(0)}  (\omega) \gg 4 \pi \chi^{(2)} (- \omega; \omega, 0) \mathcal{E}^0 .
\end{equation}

Using Eqs. \ref{eq:DielectricFunction} and \ref{eq:SmallNonlinearity}, the effective refractive index to first order in the static electric field is given by
\begin{eqnarray}\label{eq:RefractiveIndex}
n(\omega) &=& \sqrt{ \epsilon(\omega) } \nonumber \\
&=& n_0(\omega) + \frac {2 \pi \chi^{(2)} (- \omega; \omega, 0)} {n_0(\omega)} \mathcal{E}^0 ,
\end{eqnarray}
where $n_0(\omega) = \sqrt{\epsilon^{(0)}(\omega)}$ is the linear refractive index.
Eq. \ref{eq:RefractiveIndex} is sometimes written more compactly as,
\begin{equation}\label{eq:RefrativeIndexCompact}
n(\omega) = n_0(\omega) + n_1(\omega) \mathcal{E}^0 ,
\end{equation}
where $\mathcal{E}^0$ is understood to be the applied voltage divided by the distance between the electrodes.  Comparison of Eqs.~\ref{eq:RefractiveIndex} and \ref{eq:RefrativeIndexCompact} leads to the conclusion that $n_1$ can be made arbitrarily large when $n_0$ vanishes, an effect that has been used by Boyd to make ultralarge $n_2$ \cite{zahir16.01} -- the next higher-order term -- in indium tin oxide.  However, since the figure of merit is a function of competing effects that depend on the same underlying quantum parameters, the largest nonlinearity may not yield an optimal device material.

Eq. \ref{eq:RefrativeIndexCompact} shows how the refractive index can be controlled through an externally applied electric field.  Note that $n_0 $ and $n_1$ are complex quantities and that $n_0(\omega)$ contains the linear response of both the host material and the dye molecules.  As we later show, the real parts are related to the refractive index and the voltage-dependent refractive index while the imaginary parts are related to the the absorption coefficient.

The next step is to relate the susceptibilities and refractive indices, which are bulk properties, to their quantum origins.  In a single component system, such as a material made of identical noninteracting molecules, $n_0$ and $n_1$ are calculated from the same transition moments and eigenenergies, so one is a function of the other.  Since a device requires a transparent material, the imaginary part of the refractive index, $n_{0I}$ should be as small as possible so that the material is transparent; and, the real part of the electric-field-dependent refractive index coefficient should be as large as possible.  However, the interdependence between  $n_0$ and $n_1$ often makes it difficult to achieve the right balance.  For a given material, certain frequency ranges are found to be ideal while others fall short.

\subsection{The half-wave voltage $V_\pi$}
The half-wave voltage for an electro-optic device indicates the voltage required to actuate the device -- that is the voltage required to generate a phase shift of half a wavelength\cite{boyd09.01}, a Mach-Zehnder modulator as an example. The condition for determining the half-wave voltage begins with
\begin{equation}
    \text{Re}[\Delta n]L = \lambda/2
\end{equation}
where $\Delta n$ is the change in refractive index induced by the half-wave voltage and $L$ indicates the propagation length of the device. The change in refractive index is given by the field dependent part of Eq. \ref{eq:RefractiveIndex}, yielding
\begin{equation}\label{eq:HalfWaveVoltage}
    V_{\pi}=\frac{d}{L}\frac{c}{2\omega}\left(\text{Re}\left[\frac{\chi^{(2)}(-\omega;\omega,0)}{n_0(\omega)}\right]\right)^{-1},
\end{equation}
where we have taken the voltage to be $V_{\pi}=\mathcal{E}^0d$ as applied between electrodes separated by a distance $d$.

\subsection{The optical loss $\Lambda$}
In a two component system such as a dye-doped polymer, the polymer is usually of high optical quality and low optical nonlinearity, so its optical loss is low while the dye is the source of the nonlinear-optical response and adds to the loss. The device loss can be determined in terms of the ratio of the input intensity to the output intensity as
\begin{equation}
    \Lambda(\omega) = -10 \log_{10}\left(\frac{I}{I_0}\right)\text{dB},
\end{equation}
where the ratio is determined by the exponential attenuation of the electric field over the length of the device:
\begin{equation}
    \frac{I}{I_0} = \left(e^{-k\text{Im}[n_o(\omega)]L}\right)^2.
\end{equation}
Taking this to be the case, we find that the loss as a function of frequency can be expressed as
\begin{equation}\label{eq:Loss}
    \Lambda(\omega) = \frac{20}{\ln 10}\frac{\omega L}{c}\text{Im}[n_0(\omega)],
\end{equation}
in units of dB.

\subsection{Expression for the figure of merit}
The quantities in the previous sections determine the parameters needed to determine the figure of merit in terms of the real and imaginary parts of $\chi^{(2)}$ and the index of refraction as a function of frequency. We define the figure of merit to be
\begin{equation}\label{eq:FOMbulk}
    \xi = \frac{1}{\Lambda V_{\lambda/2}} = \frac{\ln 10}{10 d}\frac{\text{Re}\left[\frac{\chi^{(2)}(-\omega;\omega,0)}{n_0(\omega)}\right]}{\text{Im}[n_0(\omega)]}
\end{equation}
in units of $V^{-1}\text{dB}^{-1}$. It is interesting to note that both the real and imaginary parts of $\chi^{(2)}$ contribute to the figure of merit as
\begin{equation}
    \text{Re}\left[\frac{\chi^{(2)}}{n_0}\right] = \frac{1}{|n_0|^2}\text{Re}[\chi^{(2)}n^*_0]
        = \frac{\chi^{(2)}_\text{R} n_{0,\text{R}}+\chi^{(2)}_\text{I} n_{0,\text{I}}}{|n_0|^2},
\end{equation}
and that the real and imaginary parts of $\chi^{(2)}$ will peak at different frequencies near resonance.

\section{Relationship Between Molecular and Bulk Response}\label{sec:micro-macro}

The bulk response of an electro-optic device is directly dependent on the molecular responses of the materials which mediate the light-matter interaction. For a dye-doped polymer system of a reasonable size, the polymer can be taken as linear and lossless while the dye dopants provide the nonlinearity and linear loss. Together, the linear part of the dielectric function is given by
\begin{equation}
    \epsilon^{(0)}(\omega)= 1+4\pi\chi^{(1)}_\text{poly}+4\pi N\langle \alpha^*(-\omega;\omega) \rangle
\end{equation}
and the nonlinear dielectric function is given by
\begin{equation}
    \epsilon^{(1)}(\omega) = 4\pi N\langle\beta^*(-\omega;\omega,0)\rangle
\end{equation}
where $N$ is the number density of dye molecules, the angled brackets indicate the orientational average of dye molecules as determined by the fabrication process, and the asterisk indicates the dressed polarizability, taking into account the local fields from the surrounding media.

\subsection{Orientational Order}

The contribution to the first order susceptibility from the dye dopants is given by\cite{kuzyk98.01}
\begin{equation}\label{eq:chi1PoledPolymer}
N \left< \alpha^* \right>_{ij} =  N \int d\Omega \, a_{iI}(\vec{\Omega}) a_{jJ}(\vec{\Omega})  G(\Omega) \alpha_{IJ}^*,
\end{equation}
where $a(\vec{\Omega})$ is the Euler rotation matrix -- which is a function of the Euler angles $(\theta, \phi, \psi)$, the angular integration element is given by $ d \Omega = d \cos\theta \, d \phi \, d \psi$, and $G(\theta)$ is the orientational distribution function of the dopant molecules.  Note that summation convention applies, so the upper case indices -- which represent coordinates fixed to the dopant molecule -- are summed, and the lower case indices represent the laboratory frame.

Since we assume that all the nonlinearity comes from the host dyes, as is the case by design, then,\cite{singe87.01}
\begin{eqnarray}\label{eq:chi2PoledPolymer}
\chi_{ijk}^{(2)} & = & N \left< \beta^* \right> \nonumber \\
&=&   N \int d\Omega \, a_{iI}(\vec{\Omega}) a_{jJ}(\vec{\Omega}) a_{kK}(\vec{\Omega}) G(\Omega) \beta_{IJK}^*,
\end{eqnarray}
Recall that $\beta^*$ represents the dressed hyperpolarizability, which accounts for the surrounding material's screening or enhancement of the local fields at the molecular site.  The relationship between local field models and the dressed hyperpolarizability will be described in more detail later.

To simplify derivations for the sake of illustration, we assume that the symmetry of the material is described by one unique axis about which there is $\infty_{mm}$ symmetry so that only the polar Euler angle is relevant.  This is true for electric-field-poled dye-doped polymers, which we will use as an example.  Secondly, we assume that the molecule is one-dimensional so that the only non-vanishing component of polarizability and hyperpolarizability are $\alpha \equiv \alpha_{zz}$ and $\beta \equiv \beta_{zzz}$.  For any ordering potential $U(\theta)$ that leads to $\infty_{mm}$ symmetry -- such as an applied electric field -- the orientational distribution function is calculated using the partition function, which depends only on the polar angle $\theta$ and is of the form\cite{kuzyk98.01,singe87.01}
\begin{eqnarray}\label{eq:OrientDistribute}
G(\cos \theta) & = & \frac { \exp \left( - U(\theta) /kT \right)} {\int_{-1}^{+1} d ( \cos \theta ) \, \exp \left( - U(\theta)  \theta /kT \right)} \\
 & = & \frac { \exp \left( \mu^* \bar{\mathcal{E}} \cos \theta /kT \right)} {\int_{-1}^{+1} d ( \cos \theta ) \, \exp \left( \mu^* \bar{\mathcal{E}} \cos \theta /kT \right)} ,
\end{eqnarray}
where $\mu^*$ is the dressed dipole moment of the molecule, $\bar{\mathcal{E}}$ and $T$ are the applied electric field and temperature when the molecular orientations lock in place as the material cools. These orientational order effects are entirely dependent on the fabrication procedures and are not effected by normal device operation. A detailed description of orientational distribution functions can be found in Appendix \ref{sec:E-Field-Appendix} and other sources.\cite{singe87.01,kuzyk98.01}

\subsubsection{Linear Susceptibility}

The contribution to the first order susceptibility from the dye dopants in the 1D molecular approximation is given by
\begin{eqnarray}\label{eq:chi1Polymer1D}
N \left< \alpha^* \right>_{zz} &=&  N \int_{-1}^{+1} d ( \cos \theta ) \, \cos^2 \theta  G(\cos \theta) \alpha_{ZZ}^*  \\
&=& N \alpha_{ZZ}^*  \frac {\int_{-1}^{+1} dx \, x^2  \exp \left( a x\right) } { \int_{-1}^{+1} dx \, \exp \left( a x\right)},\label{eq:chi1PoledPolymer1D}
\end{eqnarray}
where Eq. \ref{eq:chi1Polymer1D} is the general result for any axial-only ordering potential and Eq. \ref{eq:chi1PoledPolymer1D} is for the special case of a poled polymer with $a = \mu^* \bar{\mathcal{E}} / kT$.

In the general case, it is convenient to expand the orientational distribution function as a series in the orthonormal Legendre polynomials (See Appendix \ref{sec:DistFunctionsAppendix} and Eq. \ref{eq:OrientationalDistFunc}).  According to Eq. \ref{eq:P2}, $x^2 = (2 P_2 (x) + 1 )/3$, so Eq. \ref{eq:chi1Polymer1D} can be expressed in terms of the order parameter $\left< P_2 \right>$, yielding
\begin{eqnarray}\label{eq:chi1Polymer1DEval}
N \left< \alpha^* \right>_{zz} &=&  N \alpha_{ZZ}^*  \left( \frac {2} {3} \left< P_2 \right> +  \frac {1} {3} \right)  \\
&=& N \alpha_{ZZ}^* \Bigg( 1 + 2 \left(  \frac {kT} {\mu^* \bar{\mathcal{E}}} \right)^2  \nonumber \\
 & - & \frac {2kT} {\mu^* \bar{\mathcal{E}}} \coth \left( \frac {\mu^* \bar{\mathcal{E}}} {kT} \right) \Bigg) , \label{eq:chi1PoledPolymer1DEval2}
\end{eqnarray}
and Eq. \ref{eq:chi1PoledPolymer1DEval2} is obtained using Eq. \ref{eq:<P2>(E)}.  Note that for full alignment, as one gets for an infinite applied electric field, $\left< P_2 \right> \rightarrow 1$ and $N \left< \alpha^* \right>_{zz} = N \alpha_{ZZ}^* $; i.e., the macroscopic and molecular values are the same, as expected.

\subsubsection{Second-Order Susceptibility}

The second order susceptibility originates in the dye dopants.  In the 1D molecular approximation, the second-order susceptibility is given by,
\begin{eqnarray}\label{eq:chi2Polymer1D}
\chi^{(2)}_{zzz} &=&  N \int_{-1}^{+1} d ( \cos \theta ) \, \cos^3 \theta  \, G(\cos \theta) \beta_{ZZZ}^*  \\
&=& N \beta_{ZZZ}^*  \frac {\int_{-1}^{+1} dx \, x^3  \exp \left( a x\right) } { \int_{-1}^{+1} dx \, \exp \left( a x\right)},\label{eq:chi2PoledPolymer1D}
\end{eqnarray}
where as in the linear case, Eq. \ref{eq:chi2Polymer1D} is the general result for an axial-only ordering potential and Eq. \ref{eq:chi2PoledPolymer1D} is for the special case of a poled polymer with $a = \mu^* \bar{\mathcal{E}} / kT$.

For the general case, the orientational distribution function is again a series in the orthonormal Legendre polynomials as given by Eq. \ref{eq:OrientationalDistFunc}.  According to Eq. \ref{eq:P3}, $x^3 = (2 P_3 (x) + 3 P_1 (x) )/3$, so Eq. \ref{eq:chi2Polymer1D} can be expressed in terms of the order parameters $\left< P_1 \right>$ and $\left< P_3 \right>$,
\begin{eqnarray}\label{eq:chi2Polymer1DEval}
N \left< \beta^* \right>_{zzz} &=&  N \beta_{ZZZ}^*  \left( \frac {3} {5} \left< P_1 \right> +  \frac {2} {5}  \left< P_3 \right> \right)  \\
&=& N \beta_{ZZZ}^* \left[ -   \frac {3kT} {\mu^* \bar{\mathcal{E}}}  - 6 \left(  \frac {kT} {\mu^* \bar{\mathcal{E}}} \right)^3\right. \nonumber \\
 & + & \left. \left( 1 + 6 \left( \frac {kT} {\mu^* \bar{\mathcal{E}}} \right)^2 \right) \coth \left( \frac {\mu^* \bar{\mathcal{E}}} {kT} \right) \right],\label{eq:chi1PoledPolymer1DEval}
\end{eqnarray}
where Eq. \ref{eq:chi1PoledPolymer1DEval} is obtained using Eqs. \ref{eq:<P1>(E)} and \ref{eq:<P3>(E)}.  Note that for full alignment, as one gets for an infinite applied static electric field, $\left< P_1 \right> = \left< P_3 \right> \rightarrow 1$ and $N \left< \beta^* \right>_{zzz} = N \beta_{ZZZ}^* $; i.e., the macroscopic and molecular values are the same as expected.  In the zero-field limit of a dye-doped polymer, we get,
\begin{equation}\label{eq:chi2Polymer1DEvalZeroField}
N \left< \beta^* \right>_{zzz} =  N \beta_{ZZZ}^*   \frac  {\mu^* \bar{\mathcal{E}}} {5kT} ,
\end{equation}
in agreement with the thermodynamic model of poling.\cite{singe87.01}

\begin{figure}
    \includegraphics{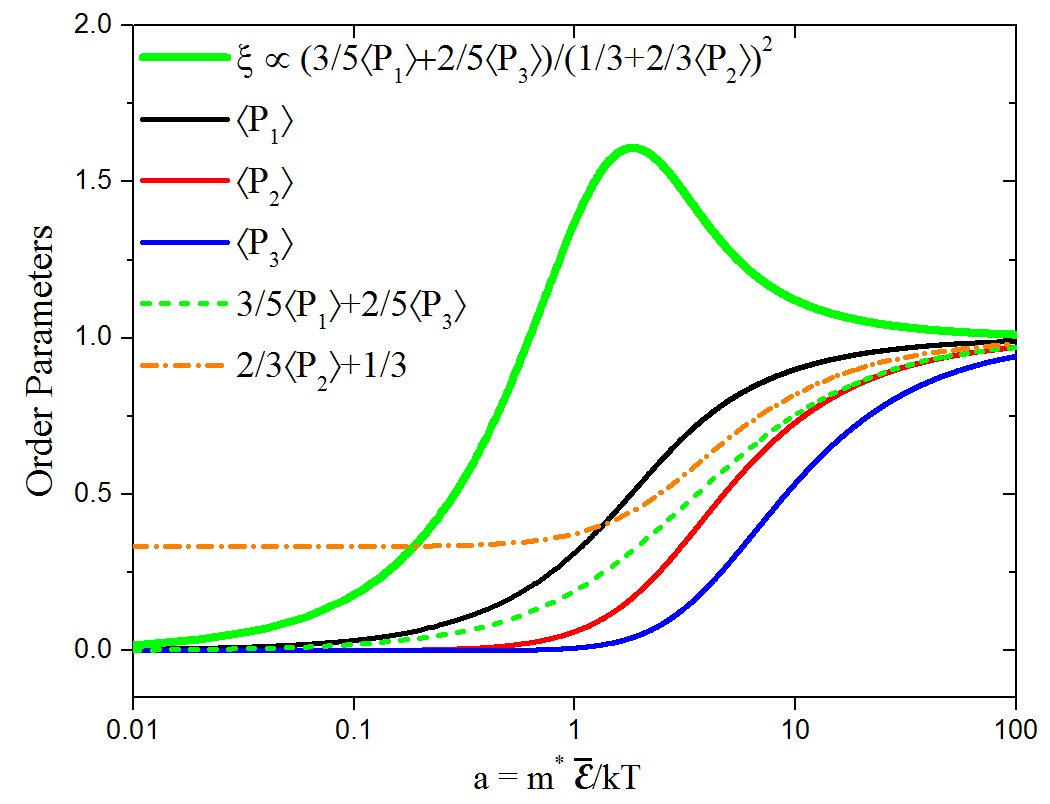}
    \caption{Various order parameters as a function of electric field strength $\bar{\mathcal{E}}$ used to align the dye dopants during fabrication. The electro-optic figure of merit is proportional to the function peaked at $m^*\bar{\mathcal{E}}/kT = 1.84$.}
    \label{fig:polingFOM}
\end{figure}

The orientational order parameters can be varied independently of the nonlinear-optical properties of the molecules using a variety of external influences during material fabrication.  Therefore, these parameters provide an avenue for material engineering.  As shown in Fig. \ref{fig:polingFOM}, each of these order parameters increase monotonically as a function of poling field, but the contribution to the electro-optic figure of merit from the order paramters described by Eq. \ref{eq:FOMbulk} is given by
\begin{equation}\label{eq:OrderParPartFOM}
\xi \propto \frac {\frac {3} {5} \left< P_1 \right> + \frac {2} {5} \left< P_3 \right>} {\frac {1} {3} + \frac {2} {3} \left< P_2 \right>}.
\end{equation}
This ratio of order parameters peaks near $m^*\bar{\mathcal{E}}/kT = 1.8$. The temperature and poling field applied to the device material during fabrication provides the means with which to optimize functionality while being independent of the details of the molecular characteristics intrinsic to the dye molecules. For simplicity of discussion, for the rest of this work we will take the orientational order coefficient to be 1, corresponding to infinite poling fields, and focus on the quantum properties of the molecules.

\subsection{Dressed Properties and Local Fields}

The vacuum polarizability $\alpha$ is related the dressed value through the fourth-rank local field tensor $L^{(1)} (- \omega; \omega)$,
\begin{equation}\label{eq:LocalF(1)-Tensor}
\alpha_{IJ}^* (- \omega; \omega) = L_{II'JJ'}^{(1)}(- \omega; \omega) \alpha_{I'J'} (- \omega; \omega),
\end{equation}
where the primed subscripts are summed.  Similarly, the vacuum hyperpolarizability $\beta$ is related to the dressed value through the sixth-rank local field tensor $L^{(2)} (- \omega_\sigma; \omega_1, \omega_2)$,
\begin{eqnarray}\label{eq:LocalF-Tensor}
\beta_{IJK}^* (- \omega_\sigma; \omega_1, \omega_2) &=& L_{II'JJ'KK'}^{(2)}(- \omega_\sigma; \omega_1, \omega_2) \nonumber \\
& \times & \beta_{I'J'K'} (- \omega_\sigma; \omega_1, \omega_2) ,
\end{eqnarray}
where the primed subscripts are summed and energy conservation demands that $- \omega_\sigma + \omega_1 + \omega_2 = 0$.

Now we are prepared to evaluate the local field model, which depends on the refractive index.  We use the simple Lorentz-Lorenz local field model,\cite{kuzyk98.01, maki91.01} which for the 1-D molecule has only one non-zero component.  For the polarizability, the only nonvanishing component for the local field correction to the polarizability is
\begin{equation}\label{eq:1D-LocalFieldAlpha}
L^{(1)} \equiv L_{ZZ^\prime ZZ^\prime}(- \omega; \omega) = \frac {n^2(\omega)+2} {3} \cdot \frac { n^2(\omega)+2} {3}
\end{equation}
and for the hyperpolarizability is
\begin{eqnarray}\label{eq:1D-LocalFieldBeta}
L^{(2)} &\equiv& L_{ZZ^\prime ZZ^\prime ZZ^\prime}(- \omega_\sigma; \omega_1, \omega_2) = \frac {n^2(\omega_\sigma)+2} {3} \nonumber \\
&\times&  \frac { n^2(\omega_1)+2} {3} \cdot\frac {n^2(\omega_2)+2} {3} ,
\end{eqnarray}
where $n (\omega)$ is the average refractive index of the composite material in the z-direction at frequency $\omega$.

The dressed polarizability in the $z$ direction of the dopant molecule for the electro-optic effect is given by
\begin{equation}\label{eq:LocalF-TensorAlpha}
\alpha^* (- \omega; \omega) = L^{(1)} (- \omega; \omega) \alpha(- \omega; \omega) ,
\end{equation}
and the dressed hyperpolarizability is given by
\begin{equation}\label{eq:LocalF-TensorBeta}
\beta^* (- \omega_\sigma; \omega_1, \omega_2) = L^{(2)}(- \omega; \omega, 0)\beta(- \omega; \omega, 0).
\end{equation}
The resulting refractive index depends on the dressed polarizability and hyperpolarizability while the local field factors depend on the average refractive index. This could be solved self-consistently by iteration, but here we will use an algebraic method by equating the dielectric function to the dressed polarizability, which contains the dielectric function, or
\begin{align}
    \epsilon(\omega) &= 1 + 4\pi\chi^{(1)}(-\omega,\omega)\nonumber\\
    &= n_\text{poly}^2+4\pi N\alpha(-\omega;\omega)\left(\frac{\epsilon(\omega)+2}{3}\right)^2
\end{align}
then solving the quadratic equation for $\epsilon(\omega)$. We have neglected the second order correction to the dielectric function in this context as its contribution to the local field factor is much smaller than the first order correction, even on resonance. The result we obtain is
\begin{equation}\label{eq:epsLFF}
    \epsilon(\omega) = \frac{9-16N\pi\alpha_\omega-3\sqrt{9-32N\pi\alpha_\omega-16N\pi\alpha_\omega n^2_\text{poly}}}{8N\pi\alpha_\omega}
\end{equation}
where the sign choice was made to require that the vacuum value of the dielectric function be given by $\epsilon(\omega) = 1$. Eq. \ref{eq:epsLFF} is then substituted into Eqs. \ref{eq:1D-LocalFieldAlpha} and \ref{eq:1D-LocalFieldBeta} to determine the local field corrections as a function of the vacuum polarizability, which can be determined from dilute gas-phase measurements.

\section{The Fundamental Limits on the Figure of Merit}\label{sec:limits}

In terms of microscopic properties, the electro-optic figure of merit Eq. \ref{eq:FOMbulk} is given by
\begin{equation}\label{eq:FOMmicro}
\xi = \frac{\ln10}{10d} \frac{\text{Re}\left[\frac{N\langle\beta^*\rangle}{n_\text{poly}+2\pi N\langle\alpha^*\rangle/n_\text{poly}}\right]}{\text{Im}[n_\text{poly}+2\pi N\langle\alpha^*\rangle/n_\text{poly}]}.
\end{equation}
It would seem that to optimize this ratio requires that the magnitude of $\beta$ be as large as possible while the imaginary part of $\alpha$ be as small as possible, all while capturing the trade-offs required by quantum characteristics.  To do so, we use an approach that is similar to to the one developed for determining the limits of $\beta$ alone.

It has been postulated, and supported by significant empirical evidence, that the optimum hyperpolarizability is obtained for a three-level system\cite{kuzyk00.01}. Adding additional states with appreciable transition strength only serves to reduce the overall response. The linear polarizability, on the other hand, is optimized for a two-level system and can be minimized for a three-level model, in principle.  So, we proceed by approximating both $\alpha$ and $\beta$ by a three-level model under the constraints of select sum rules and calculate the figure of merit as a function of a minimal number of presumedly-independent scale-invariant parameters.

\subsection{Linear polarizability}

\begin{figure}
    \includegraphics[width=\linewidth]{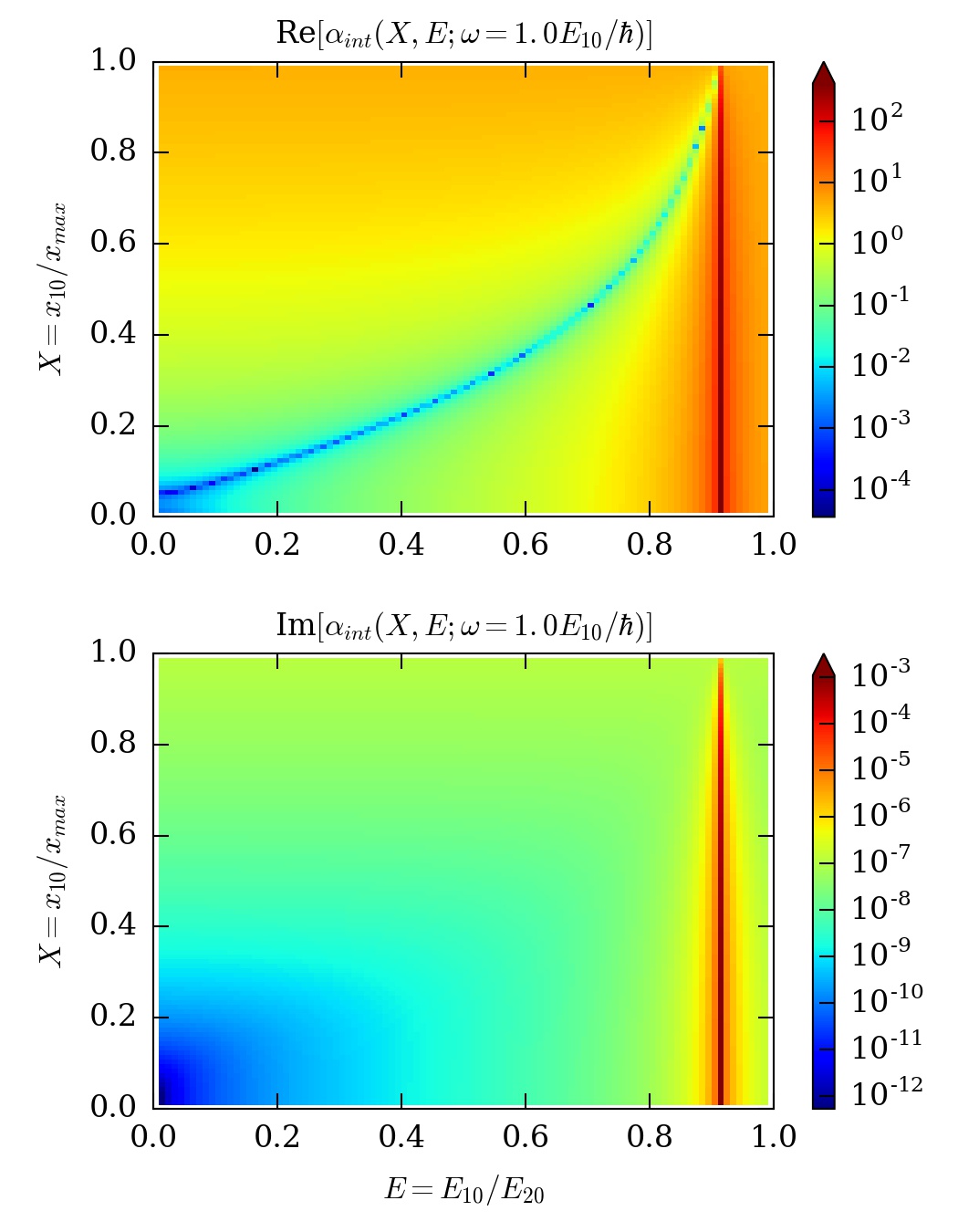}
    \caption{The real and imaginary parts of the intrinsic polarizability $\alpha^{\mbox{3L}} / \alpha_0^{\mbox{max}}$ for a three-state system constrained by the sum rules as a function of $X$ and $E$ at a frequency just above the first resonance with $\omega = 1.1 E_{10}/\hbar$.}
    \label{fig:alpha3L}
\end{figure}

The sum-over-states expression for the linear susceptibility for a one-dimensional molecule is given by\cite{orr71.01}
\begin{equation}\label{eq:alpha}
\alpha(-\omega;\omega) =  e^2 \sum_{n\neq0} \left[ \frac {\left|x_{0n}\right|^2} {E_{n0}+i\Gamma_{n0} + \hbar \omega} + \frac {\left|x_{0n}\right|^2} {E_{n0}-i\Gamma_{n0} - \hbar \omega}\right] ,
\end{equation}
where $\hbar \omega$ is the photon energy, $-e$ is the electron charge, $x_{n0}$ is the $(n,0)$ matrix element of the position operator, $E_{n0} = E_n - E_0$ is the difference between the eigenenergies of state $n$ and $0$ and is called the transition energy, and $\Gamma_{nm}$ is the phenomenological damping factor between states $n$ and $m$.

In order to capture the resonant properties of the figure of merit, we must make a reasonable approximation for the phenomenological damping factor. The minimum damping allowed by quantum mechanics is half the natural linewidth\cite{schif68.01,kuzyk06.03}, given by
\begin{equation}
    \Gamma_{nm} = \frac{1}{3}\left(\frac{E_{nm}}{\hbar c}\right)^3e^2|x_{nm}|^2 ,
    \label{eq:gamma}
\end{equation}
which will provide the best case scenario for resonant enhancement.

First, we calculate the fundamental limit for the off-resonance polarizability, where $\hbar \omega = 0$.  To do so, we need the sum rules, which must be obeyed by any molecular Hamiltonian.  They are given by,\cite{bethe77.01}
\begin{equation}
\sum_{n=0}^{\infty} \left( E_n - \frac {1} {2} \left( E_l + E_p
\right) \right) x_{ln} x_{np} = \frac {\hbar^2 N_{e}} {2m_e} \delta_{l,p},
\label{sumrule}
\end{equation}
where $m_e$ is the mass of the electron and $N_{e}$ the number of electrons. The sum, indexed by
$n$, is over all states of the system.  Eq. \ref{sumrule}
represents an infinite number of equations, one for each value of
$l$ and $p$.  As such, we refer to a particular equation using the notation $(l,p)$.

For the off-resonant limit we may take the damping to be negligible and Eq. \ref{eq:alpha} can be expressed as the inequality
\begin{equation}\label{eq:alphaInequality}
\alpha(0) =  2 e^2 \sum_n  \frac {\left|x_{0n}\right|^2} {E_{n0}}  \leq \frac {2 e^2} {E_{10}^2} \sum_n  E_{n0} \left|x_{0n}\right|^2,
\end{equation}
where all terms in the sum are positive definite, so the sum with increasing energies in the denominator cannot be larger than the sum with $E_{n0}$ replaced with $E_{10}$ since by definition, state 1 is of the lowest energy.  Finally, using the $(0,0)$ sum rule, Eq. \ref{eq:alphaInequality} yields the maximum value
\begin{equation}\label{eq:alphaMAX}
\alpha_0^{\text{max}} = \frac {e^2 \hbar^2 } {m_e} \frac {N_{e}} { E_{10}^2}.
\end{equation}

Note that Eq. \ref{eq:alphaMAX} makes the important statement that the largest possible polarizability is the one in which only the transition to the first excited state is allowed and all others vanish.  This is identically true for a harmonic oscillator with one electron.  As such, a two-level system optimizes the polarizability.  In this context, a two-level model refers only to the number of states that contribute to the polarizability.  Clearly, the harmonic oscillator has an infinite number of states.

We will use this same approach to {\em minimize} the imaginary part of the polarizability, which may require that many states contribute.  We will compromise by considering a three-level system, which has a second state that can draw away some of the oscillator strength to decrease the loss.  To eliminate biases due to the size of the system, we need to determine the dispersion of the polarizability for the three-level model in terms of the scale invariant parameters
\begin{equation}
    E = \frac{E_{10}}{E_{20}}\text{ and }X=\frac{x_{10}}{x_\text{max}} ,
\end{equation}
where $E$ and $X$ take values from 0 to 1 and where $x_\text{max}$ is determined by the two-state limit of the $(0,0)$ sum rule, which yields that largest possible transition moment
\begin{equation}
    x_\text{max}^2 =\frac{N_e\hbar^2}{2m_eE_{10}}.
\end{equation}

Let's begin by considering the $(0,0)$ sum rule when three states dominate, or
\begin{equation}\label{eq:3Lsumrule}
\left| x_{02} \right|^2 = E \left( x_\text{max}^2 - \left| x_{01} \right|^2 \right).
\end{equation}
The three-state model of the polarizability from Eq. \ref{eq:alpha} is given by
\begin{align}
        \alpha^\text{3L}&(-\omega;\omega) = e^2\left[|x_{10}|^2\left(\frac{1}{E_{10}-i\Gamma_{10}-\hbar\omega}        +\frac{1}{E_{10}+i\Gamma_{10}+\hbar\omega}\right)\right.\nonumber\\
        &+\left.|x_{20}|^2\left(\frac{1}{E_{20}-i\Gamma_{20}-\hbar\omega}+ \frac{1}{E_{20}+i\Gamma_{20}+\hbar\omega}\right)\right],
        \label{eq:alpha3L}
\end{align}
which using Eq. \ref{eq:3Lsumrule} becomes
\begin{align}
        \alpha^\text{3L}&(-\omega;\omega) = \frac{\alpha_0^\text{max}}{2}\left[X^2\left(\frac{1}{1-i\gamma_{10}-\tilde{\omega}}+\frac{1}{1+i\gamma_{10}+\tilde{\omega}}\right)\right.\nonumber\\
        &\left.+E(1-X^2)\left(\frac{1}{E^{-1}-i\gamma_{20}-\tilde{\omega}}+\frac{1}{E^{-1}+i\gamma_{20}+\tilde{\omega}}\right)\right],
        \label{eq:alpha3L-2}
\end{align}
where $\tilde{\omega}=\hbar\omega/E_{10}$,
\begin{equation}
    \gamma_{10}=\frac{\Gamma_{10}}{E_{10}} = \frac{N_e\alpha_\text{FS}}{6}X^2\frac{E_{10}}{m_ec^2},
    \label{eq:gamma10}
\end{equation}
and
\begin{equation}
    \gamma_{20} = \frac{\Gamma_{20}}{E_{10}} = \frac{N_e\alpha_\text{FS}}{6E^2}(1-X^2)\frac{E_{10}}{m_ec^2},
    \label{eq:gamma20}
\end{equation}
where $\alpha_\text{FS}$ is the fine structure constant. While it is natural to describe the optical frequencies in units of the lowest molecular resonance frequency, the natural linewidths cannot be expressed in this way because the ratio $E_{10}/m_e c^2$ defines yet another dimensionless quantity that remains in Equations \ref{eq:gamma10} and \ref{eq:gamma20}.  In other words, Equations \ref{eq:gamma10} and \ref{eq:gamma20} cannot be expressed in terms of only $X$ and $E$ but contain another dimensionless parameter that is scaled by the rest energy of the electron.

Figure \ref{fig:alpha3L} shows a color map of the magnitude of the real and imaginary parts of the intrinsic polarizability $\alpha^{\mbox{3L}} / \alpha_0^{\mbox{max}}$ as a function of $E$ and $X$ just above the first resonance when $\omega = 1.1 E_{10}$.  The spike near $E=0.9$ corresponds to a second excited state energy that matches the photon energy, so is the resonant response.  The blue region is far off resonance, where the imaginary part of the polarizability is minimum.  The minimum curve in the real part of the polarizability can be attributed to the opposite signs of the first and second state contributions, where cancelation requires a specific set of transition moments.

\subsection{First hyperpolarizability}

\begin{figure}
    \includegraphics[width=\linewidth]{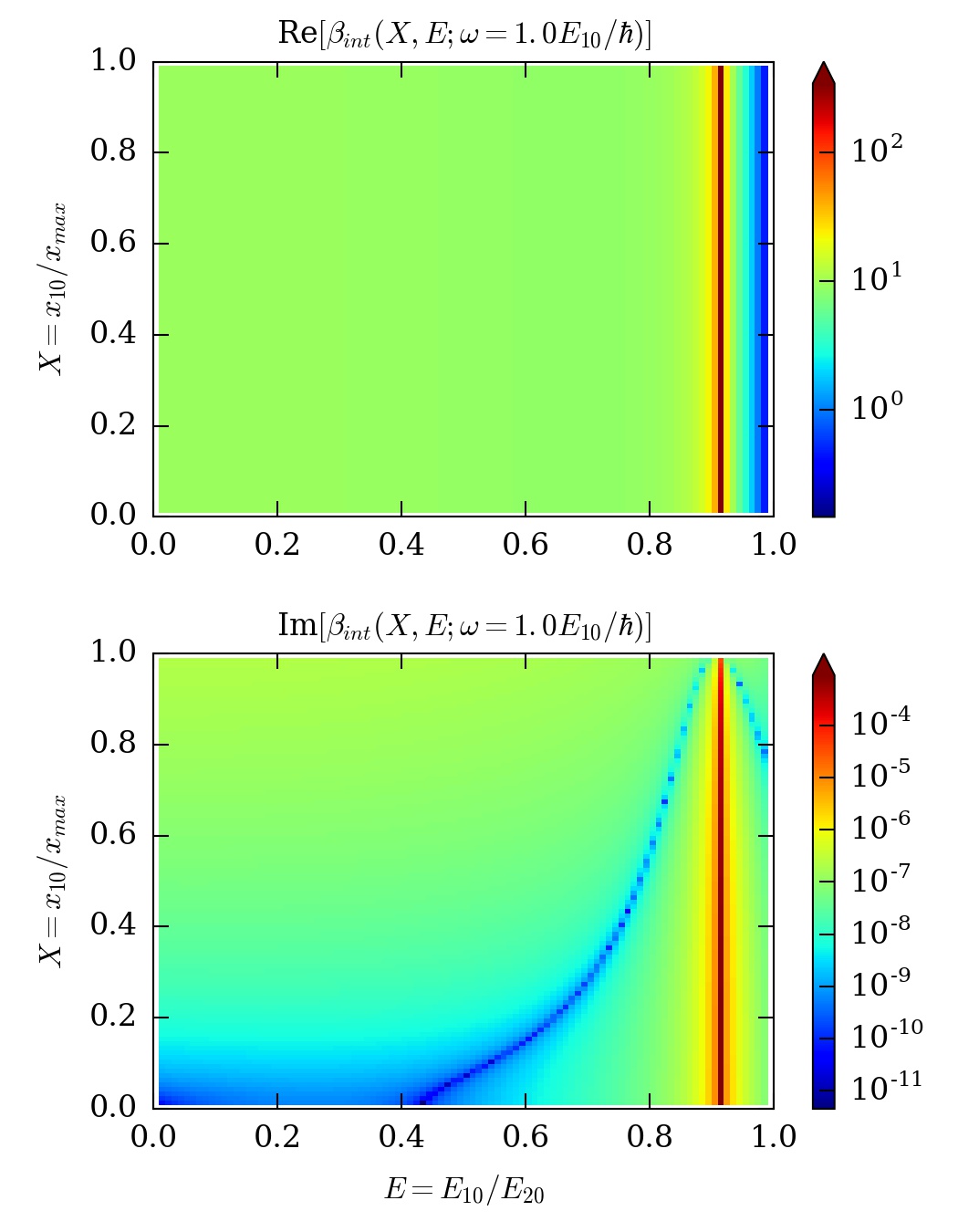}
    \caption{The real and imaginary parts of the intrinsic hyperpolarizability $\beta^{\mbox{3L}} / \beta_0^{\mbox{max}}$ as a function of $X$ and $E$ for a three-level model constrained by the sum rules at a frequency just above the first resonance with $\omega = 1.1 E_{10}/\hbar$.}
    \label{fig:beta3L}
\end{figure}

It is worthwhile to step back and review the approach in calculating fundamental limits.  In the derivation of Eq. \ref{eq:alphaMAX}, we found that the limit of the polarizability is characterized by a molecule with a transition from the ground state to only one excited state.  One might then expect that the same might be true for hyperpolarizability.  It is simple to show that any system with only one non-zero transition from the ground state that obeys the sum rules must have a vanishing hyperpolarizability.  As such, it was proposed that the limit is characterized by a system with two excited states rather than one.\cite{kuzyk00.01}  This guess, which has never been proven but appears to always hold, is called the three-level ansatz.  In particular, it states that the SOS expression for the hyperpolarizability of a quantum system is dominated by contributions from only two excited states at the fundamental limit.\cite{kuzyk13.01}

The fundamental limit of the off-resonant hyperpolarizability is given by \cite{kuzyk00.01}
\begin{equation}\label{beta-max-off-res}
\beta_0^\text{max} = \sqrt[4]{3} \left( \frac {e \hbar} {\sqrt{m_e}} \right)^3 \frac {N_{e}^{3/2}} {E_{10}^{7/2}} .
\end{equation}
This limit is calculated by optimizing the three-level expression for the hyperpolarizability under the constraints of the sum rules, yielding
\begin{equation}\label{eq:3L-Ansatz}
\beta(X, E) = \beta_0^\text{max} f(E) G(X) ,
\end{equation}
where
\begin{equation}\label{defineGx}
G(X) = \sqrt[4]{3} \sqrt{\frac
{3} {2}} X \sqrt{1 - X^4}
\end{equation}
and
\begin{equation}\label{definefE}
f(E) = \frac {1} {2} \left(1-E \right)^{3/2} \left( 2 + 3E + 2E^2 \right) .
\end{equation}
Thus, the function to be optimized decouples in one of $E$ and one of $X$, making it a straightforward matter to find the maximum values of each function, $f(E)$ and $G(X)$.

The dispersion of the first hyperpolarizability\cite{kuzyk06.03} is not quite as simple.  The sum-rule constrained three-level model is given by
\begin{equation}
    \beta^\text{3L}(-\omega;\omega_1,\omega_2) = \beta_0^\text{max}\frac{1}{6}\frac{EE_{10}^2}{\sqrt{1-E}}D^{\text{3L}}(\omega_1,\omega_2)G(X)
    \label{eq:betamaxdisp}
\end{equation}
where
\begin{align}
    D^{\text{3L}}(\omega_1,\omega_2) = P_{\omega_1,\omega_2}&\left[D_{12}(\omega_1,\omega_2)-\frac{2E^{-1}-1}{D^{-1}_{11}(\omega_1,\omega_2)}\right.\nonumber\\\
    &\left.+D_{21}(\omega_1,\omega_2)-\frac{2E-1}{D^{-1}_{22}(\omega_1,\omega_2)}\right].
\end{align}
$E = E_{10}/E_{20}$, $X = x_{10}/x_\text{max}$, and the permutation operator $P_{\omega_1,\omega_2}$ results in the dispersion denominator
\begin{align}
    P_{\omega_1,\omega_2}&\left[D_{nm}(\omega_1,\omega_2)\right] \nonumber\\
    &=\frac{1}{2}\bigg(\frac{1}{(E_{n0}+i\Gamma_{n0}-\hbar\omega_1-\hbar\omega_2)(E_{m0}+i\Gamma_{m0}-\hbar\omega_1)}\nonumber\\ &+\frac{1}{(E_{n0}-i\Gamma_{n0}+\hbar\omega_2)(E_{m0}+i\Gamma_{m0}-\hbar\omega_1)}\nonumber\\
    &+\frac{1}{(E_{n0}-i\Gamma_{n0}+\hbar\omega_2)(E_{m0}+i\Gamma_{m0}+\hbar\omega_1+\hbar\omega_2)}\nonumber\\
    &+ \omega_1 \leftrightarrow \omega_2\ \text{for the three previous terms}\bigg).
\end{align}
In Eq. \ref{eq:betamaxdisp}, the factor multiplying $\beta_0^{\mbox{max}}$ is dimensionless and describes the effects of dispersion on the limits.

Figure \ref{fig:beta3L} shows a color map of the sum-rule constrained three-level model of the intrinsic hyperpolarizability $\beta^{\mbox{3L}} / \beta_0^{\mbox{max}}$ as a function of $E$ and $X$.  As in the case of the polarizability, resonance features are observed with peaks.  Traditionally, the focus of material design would be on maximizing the off-resonant real part of the hyperpolarizability, since that has the lowest loss.  Then the figure of merit would be evaluated for materials designed in this way.  However, the best figure of merit might be in a region where both the loss and the nonlinearity are low, as long as the loss drops more dramatically than the nonlinearity, as we see in the $(E,X) \rightarrow (0,0)$ domain or in areas where the loss is high but the nonlinearity is even higher.  As we show below, the best material design requires that the figure of merit be optimized directly rather than focusing on one particular contribution.

\section{The Dispersion of the Figure of Merit}\label{sec:fomdispersion}

As is evident from Eq. \ref{eq:FOMbulk}, the figure of merit does not decouple in such a way as to allow definitive analysis of the individual molecular characteristics that contribute to it.  Substituting all the pieces we calculated above into Eq. \ref{eq:FOMbulk}, the three-level ansatz indicates that the of the figure of merit can be expressed as
\begin{align}
    \xi &= \frac{\ln10}{10d} \frac{\text{Re}\left[\frac{N\langle\beta^*\rangle}{n_\text{poly}+2\pi N\langle\alpha^*\rangle/n_\text{poly}}\right]}{\text{Im}[n_\text{poly}+2\pi N\langle\alpha^*\rangle/n_\text{poly}]}\\
    &\approx \frac{\ln10}{40\pi^2d} \frac{n_\text{poly}^2\beta_0^\text{max}}{N(\alpha_0^\text{max})^2} \frac{\text{Re}\left[\frac{L^{(2)}\beta_\text{int}^\text{3L}(-\omega;\omega,0)} {\tilde{n}+L^{(1)}\alpha_\text{int}^\text{3L}(-\omega;\omega)}\right]} {\text{Im}[\tilde{n}+L^{(1)}\alpha_\text{int}^\text{3L}(-\omega;\omega)]},
    \label{eq:FOMmax}
\end{align}
where $\tilde{n} = n_\text{poly}^2/2\pi\alpha^\text{max}_0N$ is a real, scale invariant quantity, which only depends on the polymer refractive index and volume fraction of dye molecules ($\alpha^\text{max}_0N$ is approximately the volume fraction of dye molecules).  Significant evidence supports the three-level ansatz, which states that the nonlinear-optical response of a molecule near the fundamental limit (or even at a local maximum) is well approximated by a three-state model.  If true, Eq. \ref{eq:FOMmax} can be viewed as an upper bound of the figure of merit for a given value of $E$ and $X$ by virtue of the hyperpolarizability in the numerator.

\begin{figure*}
    \includegraphics[width=\linewidth]{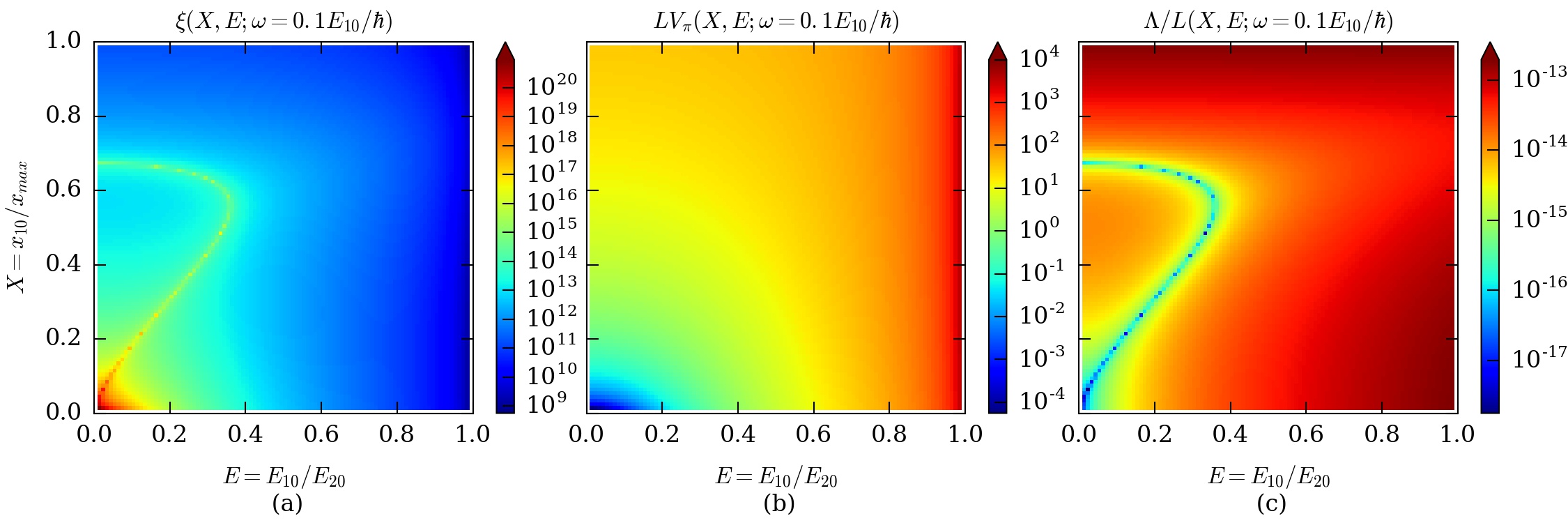}
    \caption{(a) A logarithmic plot of the figure of merit given by Eq. \ref{eq:FOMmax} in the zero-optical-frequecny limit in units of $V^{-1} \cdot dB^{-1}$ and (b) The half-wave voltage/length product in volt$\cdot$cm as a function of $X$ and $E$. (c) The loss per unit length in $dB/cm$ on a linear scale.}
    \label{fig:fom_Vpi_Loss_w0.1}
\end{figure*}

\begin{figure*}
    \includegraphics[width=\linewidth]{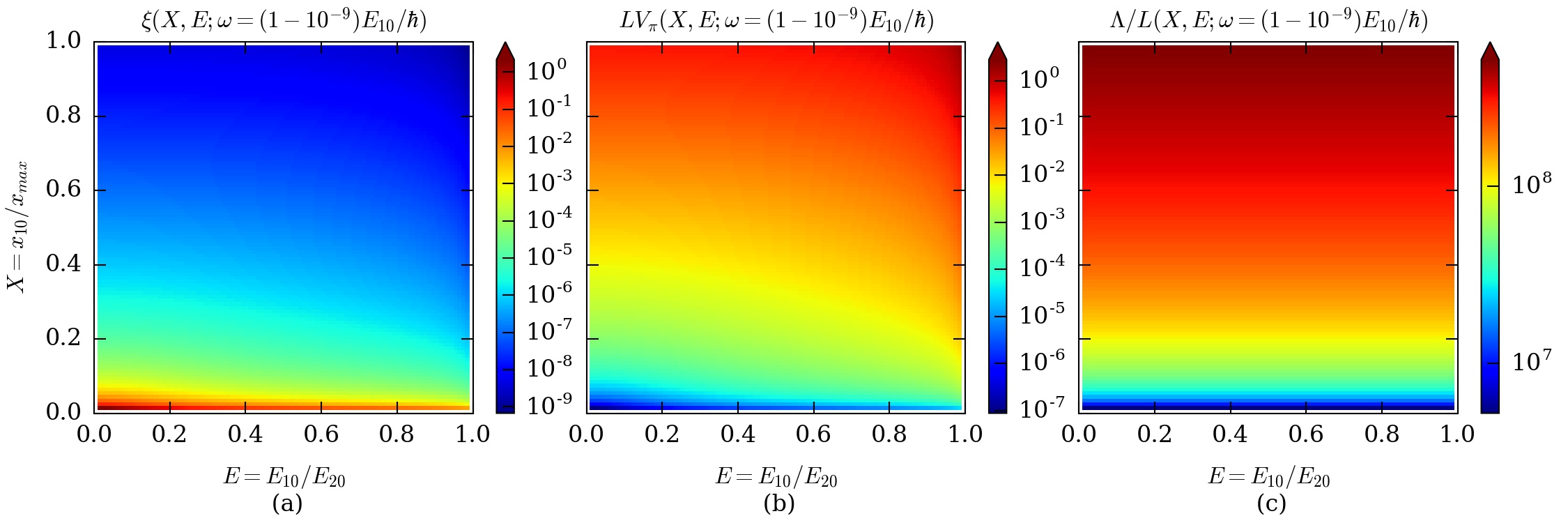}
    \caption{(a) The figure of merit given by Eq. \ref{eq:FOMmax} in units of $V^{-1} \cdot dB^{-1}$ as a function of $X$ and $E$ at a frequency approximately one natural linewidth below resonance. (b) The half-wave voltage and length product in volt$\cdot$cm. (c) The loss per unit length in $dB/cm$.}
    \label{fig:fom_Vpi_Loss_NearRes}
\end{figure*}

\begin{figure*}
    \includegraphics[width=\linewidth]{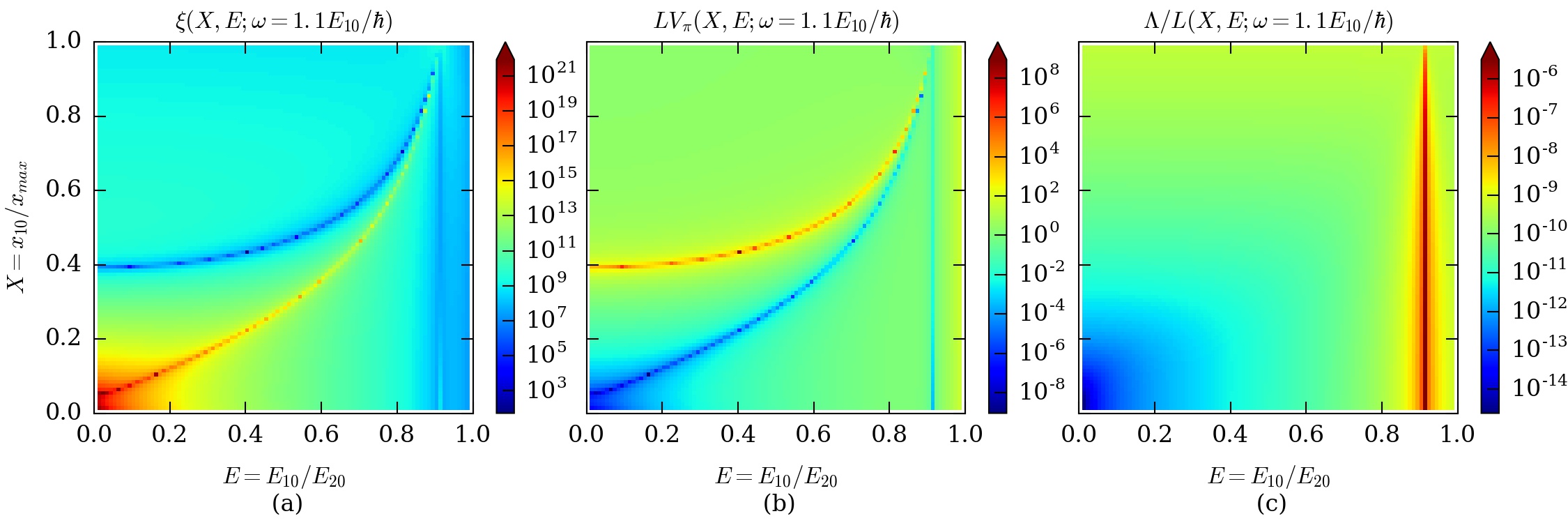}
    \caption{(a) The figure of merit given by Eq. \ref{eq:FOMmax} in units of $V^{-1} \cdot dB^{-1}$ as a function of $X$ and $E$ at a frequency just above resonance. (b) The half-wave voltage and length product in volt$\cdot$cm. (c) The loss per unit length in $dB/cm$.}
    \label{fig:fom_Vpi_Loss}
\end{figure*}

Inserting numerous equations from earlier in this work, the figure of merit can be expressed as a function of the optical frequency $\omega$, the scale invariant molecular parameters $E$ and $X$, the fabrication parameters $N$ and $\mu^*\bar{\mathcal{E}}/kT$, and the length scale set by the energy difference $E_{10}$. Figs. \ref{fig:fom_Vpi_Loss_w0.1}-\ref{fig:fom_Vpi_Loss} show the figure of merit, the product of device length and half-wave voltage, and the material loss as a function of $E$ and $X$ -- the transition moment and energy scale parameters -- for the dye-doped polymer.

To explore the character of the figure of merit for physically-reasonable parameters, we take the energy difference $E_{10} = 1 \, eV$, the number or participating electrons to be $N_e = 1$, the dopant number density $N = 10^{-5} \text{\AA}^{-3}$, the poling order parameters to be unity, and the host polymer index of refraction to be $n_\text{poly}=1.49$.  Note that for these values, $\alpha_0^{\text{max}} \approx 110 \text{\AA}^3$.  Finally, we consider a range of optical frequencies between $\omega = 0$ and $\omega = 1.4E_{10}/\hbar$. For frequencies beyond the first resonance, the second resonance will appear on the parameter space plots and we must be cautious of the higher-energy resonances which we are neglecting.

Fig. \ref{fig:fom_Vpi_Loss_w0.1}a shows the figure of merit off resonance, where the photon energy is small compared with the first excited-state energy of the dopant molecules.  A device with a switching voltage of at most 1V and a loss less than 1dB requires a figure of merit greater than unity.  The Figure of merit near $X=0$ and $E=0$ is exceptionally large, exceeding $10^{22}$, implying an infinitesimal loss and switching voltage.  Because of the low loss, the nonlinearity need not be large.  In fact, the meteoric increase of the figure of merit comes from the fact that the loss gets smaller more rapidly than the nonlinearity gets larger.

For a figure of merit of $10^{22}$ near $(X,E) = (0,0)$, as shown in \ref{fig:fom_Vpi_Loss_w0.1}b, $L V_\pi = 10^{-5}$.  Thus, the device length would need to be at least $10^{-5}$ cm for $V_{\pi} = 1$ V.  Furthermore, as seen in Figure \ref{fig:fom_Vpi_Loss_w0.1}c, the loss for this device would be $10^{-17}$ dB.  These numbers are many orders of magnitude better than is observed for any device ever demonstrated.  This is not surprising given that our calculations are setting the upper bound.  Nonlinearities are usually much smaller and the losses much higher due to inhomogeneous broadening.  Thus, decreasing the linewidth is a potentially fruitful new avenue of research for increasing the figure of merit.  The upper limits presented here suggest that much better devices are possible if the figure of merit is the target rather than first identifying large hyperpolarizability molecules and subsequently attempting to decrease their loss.

Materials with $X=0$ and $E=0$ correspond to three-level systems with a nearly degenerate ground state and no transition strength between the two lowest states. These systems would then require more states to be adequately described. As such, it might be impossible to attain such high figures of merit off resonance given that the best materials have $X \approx 0.8$ and $E > 0.5$.\cite{tripa04.01,tripa06.01}

Near resonance, the figure of merit is above unity for only a small portion of the domain of possibilities. Fig. \ref{fig:fom_Vpi_Loss_NearRes} shows the device properties for an optical frequency approximately one natural linewidth away from resonance. The figure of merit is worse than off resonance because loss grows more rapidly than the hyperpolarizability.

Beyond the first resonance, the figure of merit gets even better than it was off resonance. Fig. \ref{fig:fom_Vpi_Loss} shows the device properties in the anomalous dispersion regime.\cite{Kowal95.01}  The orange curve in Figure Fig. \ref{fig:fom_Vpi_Loss}a shows the largest figure of merit, which corresponds to the blue curve in Fig. \ref{fig:fom_Vpi_Loss}b, where $L V_{\pi}$ is at its minimum.  Thus, devices with ultrahigh performance would result along these curves that far exceed any materials demonstration to date.  Of significance is the fact that the curve of maximum figure of merit cuts across the physically-observed regime of $X \approx 0.8$ and $E > 0.5$.

Fig. \ref{fig:fom_Vpi_Loss} also shows a large portion of the parameter space for large $X$ and $E<0.91$ where the figure of merit is uniformly on the order of $10^{14}$, corresponding to devices with half-wave voltages on the order of 0.00001 Vcm and loss on the order of $10^{-11}$ dB/cm. Therefore, devices with dye molecules that can be well described by a three-level model with these parameters could, in principle, produce exemplary devices.

\begin{figure*}
    \includegraphics[width=\linewidth]{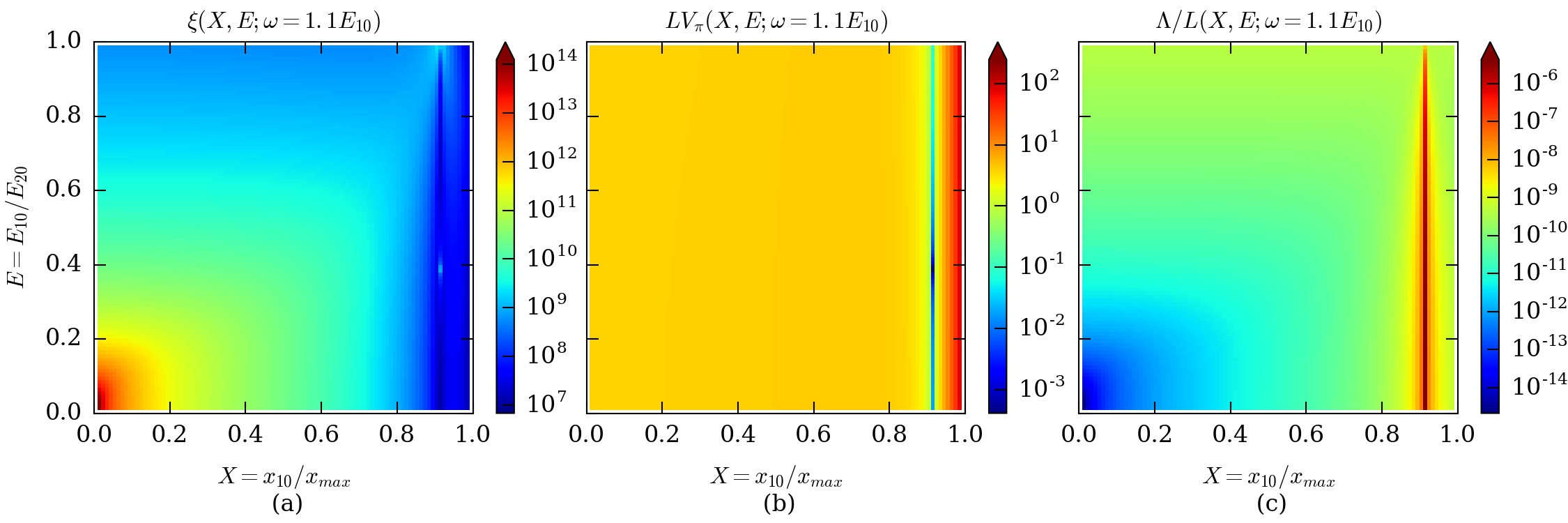}
    \caption{The material properties neglecting the local field corrections. (a) The figure of merit given by Eq. \ref{eq:FOMmax} in units of $V^{-1} \cdot dB^{-1}$ as a function of $X$ and $E$ at a frequency just above the first resonance. (b) The half-wave voltage and length product in volt$\cdot$cm and (c) the loss per unit length in $dB/cm$.}
    \label{fig:noLocFields}
\end{figure*}

The local field corrections described by Eqs. \ref{eq:1D-LocalFieldAlpha} and \ref{eq:1D-LocalFieldBeta} contribute a significant enhancement to the figure of merit overall, as well as extending the region of tolerable linear loss.  Fig. \ref{fig:noLocFields} shows how the device parameters appear without the local field corrections, to be compared with Fig. \ref{fig:fom_Vpi_Loss}.

We note that the results present here are but a small fraction of the data generated by this work.  In addition to the frequency dependence, changes in the concentration can enhance nonlinear interactions between molecules, leading to new domains that can potentially have ultra-large nonlinear figure of merit that goes well beyond the numbers calculated here.

These results together highlight four items of importance that do not appear to be appreciated in materials development:

\begin{itemize}

\item Design of electro-optic device materials requires that the figure of merit be optimized rather than the hyperpolarizability alone, an approach that is not common in the literature aside from retrospective studies of materials.  The design of other types of devices and of higher-order nonlinearity would benefit from the same approach.

\item Tuning a material to a very specific regime of anomalous dispersion might be an avenue for enhancing the figure of merit of materials that are not so remarkable off resonance.  Alternatively, materials can be designed to have the ideal dispersion in the spectral region required of an application.

\item Suppressing inhomogeneous broadening -- by for example cooling the material and/or processing it to decrease material inhomogeneity, etc. -- to bring the damping factor down to the natural linewidth can lead to enhancements of the figure of merit.

\item The local electric field factor is a critical factor in the enhancement of the figure of merit due to its simultaneous effects of increasing the nonlinear response and decreasing the loss.  Applying self-consistent fields in these calculations are of paramount importance.  This is an avenue that has seen only limited research efforts to improve materials.

\end{itemize}

\section{Conclusion}

We determined limits on the figure of merit of an electro-optic device and showed that the optimum operating configuration is either off-resonance or slightly above the first molecular resonance.  In the anomalous dispersion regime, we find the exciting prospect of a half-wave voltage on the order of 0.01 V with $10^{-4}$ dB loss for a 0.01 mm-long device for energy spectra and transition moments in the range commonly observed for organic molecules.  While these are upper limits, there is no reason why real materials cannot come near.  Even if real material fall three orders of magnitude short of this limit, one can imagine 0.1 V switching voltages in a 1 cm-long device.  However, in some ways, our results are pessimistic because they include only one electron per molecule.  Given that the nonlinearity grows more quickly with the number of electrons than does the loss, larger molecules with the correct scaling properties can have much better figures of merit.

We find that the necessary half-wave voltage and loss can be simultaneously minimized within an accessible region of molecular parameter space. These regions of optimization are not necessarily the same regions which optimize the nonlinearity, as we see that resonant features also maximize the linear absorption, and therefore result in an exceptionally lossy device. The optimum configuration differs from that of $\chi^{(1)}$ or $\chi^{(2)}$ separately suggesting a new, holistic  paradigm for materials development. Devices operating just above the first molecular resonance with a strong oscillator strength between the ground and first excited state allow for a maximal electro-optic device figure of merit.

We also find that the local electric fields, which must be determined self-consistently, play an import role in the figure of merit.  In addition, it is best when inhomogeneous broadening is minimized so that the linewidth is determined by the natural linewidth.

In general, scaling arguments can be used to investigate the properties that are required to optimize a material for a particular device application.  It may not be necessary to make molecules with large hyperpolarizabilities if other parameters such as local field factors and linewidth can be tuned.  As additional criteria are brought into the mix, the figure of merit will need to be generalized.  Undoubtedly, the requirements will change.  However, using scaling argument and limits can play an important role in optimizing the design of materials for a given application.  For the case presented here, the numbers are staggering, showing that highly-efficient electro-optic devices are possible.  Other frequency domains and concentrations may result in even a more favorable figure of merit.  Studies of this sort are underway.

We acknowledge the National Science Foundation(ECCS-1128076) for generously supporting this work.

\appendix

\section{Useful Integrals}\label{sec:UsefulIntergralsAppendix}

There are several integrals that are used often when determining order parameters, so we tabulate them here.  First,
\begin{equation}\label{eq:exponential}
\int_{-1}^{+1} dx \, \exp( a x ) = \frac {2} {a} \sinh a .
\end{equation}
From this, we can easily calculate the rest,
\begin{eqnarray}\label{eq:X.exponential}
\int_{-1}^{+1} dx \, x \exp( a x ) & = & \frac {\partial} {\partial a} \int_{-1}^{+1} dx \, \exp( a x ) \nonumber \\
& = & \frac {2} {a} \cosh a - \frac {2} {a^2} \sinh a ,
\end{eqnarray}
\begin{eqnarray}\label{eq:XX.exponential}
\int_{-1}^{+1} dx \, x^2 \exp( a x ) & = & \frac {\partial} {\partial a} \int_{-1}^{+1} dx \, x \exp( a x ) \nonumber \\
& = & \frac {2} {a} \left( 1 + \frac {2} {a^2} \right) \sinh a - \frac {4} {a^2} \cosh a , \nonumber \\
\end{eqnarray}
\begin{eqnarray}\label{eq:XXX.exponential}
\int_{-1}^{+1} dx \, x^3 \exp( a x ) & = & 2 \left[ \left( \frac {1} {a} +  \frac {6} {a^3} \right) \cosh a \right. \nonumber \\
 & - & \left. 3 \left( \frac {1} {a^2}  + \frac {2} {a^4} \right) \sinh a \right]
\end{eqnarray}

\section{Orientational Distribution Functions}\label{sec:DistFunctionsAppendix}

The orientational order of a material with $\infty_{mm}$ symmetry (i.e. symmetric under rotations about $z$, for example, which is obtained when a material is aligned with an electric field) is often described by the set of order parameters $\left< P_n \right>$, which are the coefficients in the expansion of the orientational distribution function $G(\cos \theta)$ in terms of the the orthogonal Legendre polynomials $P_n(\cos \theta)$
\begin{equation}\label{eq:OrientationalDistFunc}
G(\cos \theta) = \sum_{n=0}^\infty \frac {2n+1} {2} \left< P_n \right> P_n(\cos \theta),
\end{equation}
where $\theta$ is the polar angle, i.e. the angle measured from the symmetry.  The orthonormality condition is given by,
\begin{equation}\label{eq:LegendreOrtho}
\frac {2n+1} {2} \int_{-1}^{+1} d ( \cos \theta ) \, P_n(\cos \theta) P_m(\cos \theta) = \delta_{n,m}.
\end{equation}

\begin{figure}
    \includegraphics{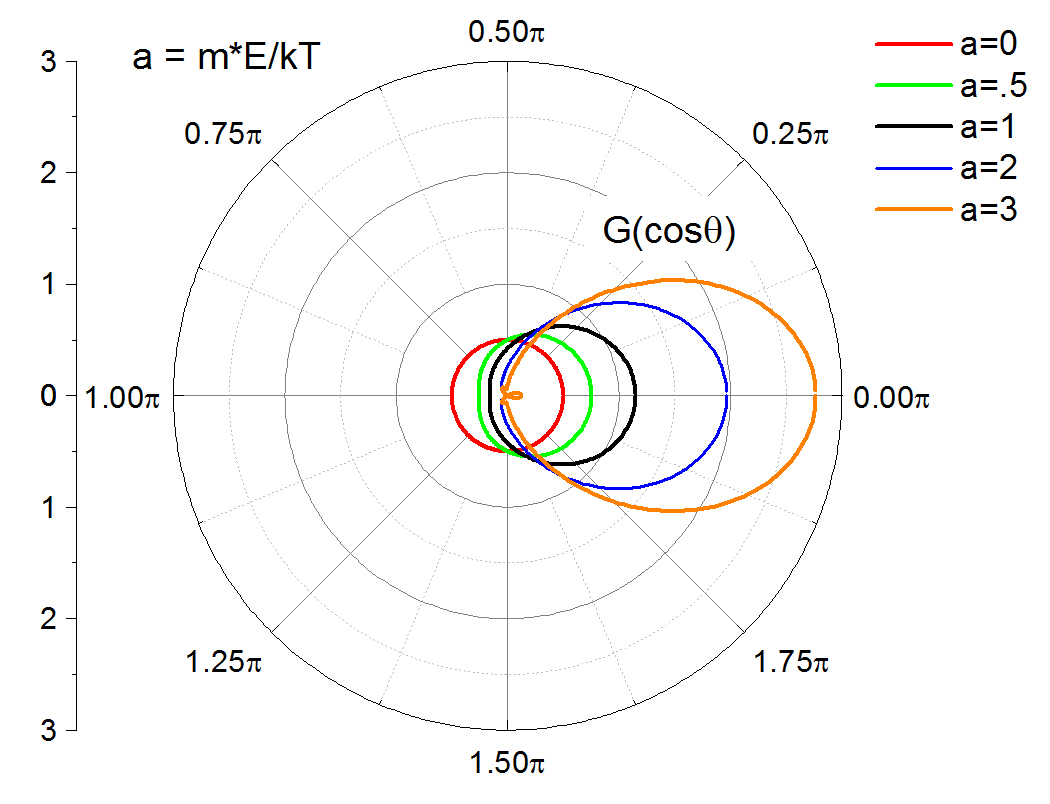}
    \caption{The first four Legendre polynomials.}\label{fig:LegendrePoly}
\end{figure}

We note that any set of orthogonal functions can be used to express the the orientational distribution function but the Legendre Polynomials are the most convenient because the nonlinear susceptibilities are related to them in a simple way.  The first five Legendre Polynomials are given by,
\begin{equation}\label{eq:P0}
P_0 (x) = 1,
\end{equation}
\begin{equation}\label{eq:P1}
P_1 (x) = x,
\end{equation}
\begin{equation}\label{eq:P2}
P_2 (x) = \frac {3 x^2 - 1} {2} ,
\end{equation}
\begin{equation}\label{eq:P3}
P_3 (x) = \frac {5 x^3 - 3x} {2} ,
\end{equation}
and
\begin{equation}\label{eq:P4}
P_4 (x) = \frac {35 x^4 - 30 x^2 + 3} {8} .
\end{equation}
Figure \ref{fig:LegendrePoly} Shows a polar plot of the first four Legendre Polynomials.

Note that we can invert the Legendre Polynomials to solve for any power of $x$ (see \url{mathworld.wolfram.com/LegendrePolynomial.html}),
\begin{equation}\label{eq:P-invert}
x^n = \sum_{\ell = n, n-2, \dots} \frac {(2 \ell + 1) n! } { 2^{(n - \ell)/2 } \left( \frac {n - \ell} {2} \right)! \left( \ell + n + 1 \right)!!} P_{\ell}(x).
\end{equation}
Another useful formula is the expansion,
\begin{equation}\label{eq:coth-series}
\coth(a) = \frac {1} {a} + \frac {a} {3} - \frac {a^3} {45} + \frac {2 a^5} {945} - \frac {a^7} {4725}  + \dots,
\end{equation}
where $a \ll 1$.

\section{Electric-Field-Induced Orientational Order}\label{sec:E-Field-Appendix}

When an electric field, $\bar{\mathcal{E}}$ is applied to a free dipole of moment $\mu^*$ that is in equilibrium at temperature $T$, the resulting order parameter $\left< P_1 \right>$ is given by,
\begin{equation}\label{eq:<P1>}
\left< P_1 \right> = \frac {\int_{-1}^{+1} d ( \cos \theta ) \, \cos \theta \exp[\mu^* \bar{\mathcal{E}} \cos \theta / kT]} {\int_{-1}^{+1} d ( \cos \theta ) \, \exp[\mu^* \bar{\mathcal{E}} \cos \theta / kT]},
\end{equation}
where $\theta$ is the angle between the dipole moment and the applied electric field.

The denominator of Eq. \ref{eq:<P1>} can be evaluated using Eq. \ref{eq:exponential} and the numerator with Eq. \ref{eq:X.exponential}, yielding,
\begin{equation}\label{eq:<P1>(E)}
\left< P_1 \right> = \coth a  - \frac {1} {a},
\end{equation}
where $a =  \mu^* \bar{\mathcal{E}} / kT$.  The limiting case of small electric field relative to thermal energies yields,
\begin{equation}\label{eq:<P1>(E)limit->0}
\lim_{a \rightarrow 0} \left< P_1 \right> = a = \frac {\mu^* \bar{\mathcal{E}}} {kT},
\end{equation}
and for large electric field yields relative to thermal energies yeilds,
\begin{equation}\label{eq:<P1>(E)limit->infinity}
\lim_{a \rightarrow \infty} \left< P_1 \right> = 1 - \frac {1} {a} = 1 - \frac {kT} {\mu^* \bar{\mathcal{E}}}.
\end{equation}

Similarly, the order parameter $\left< P_2 \right>$ is given by,
\begin{equation}\label{eq:<P2>}
\left< P_2 \right> = \frac {\int_{-1}^{+1} d ( \cos \theta ) \, \left(\frac {3 \cos^2 \theta - 1} {2} \right) \exp[\mu^* \bar{\mathcal{E}} \cos \theta / kT]} {\int_{-1}^{+1} d ( \cos \theta ) \, \exp[\mu^* \bar{\mathcal{E}} \cos \theta / kT]}.
\end{equation}
Using Eqs. \ref{eq:exponential} and \ref{eq:XX.exponential} to evaluate Eq. \ref{eq:<P2>} yields,
\begin{equation}\label{eq:<P2>(E)}
\left< P_2 \right> = 1 + \frac {3} {a^2} - \frac {3} {a} \coth a .
\end{equation}
The limiting case of small electric field relative to thermal energies yields,
\begin{equation}\label{eq:<P2>(E)limit->0}
\lim_{a \rightarrow 0} \left< P_2 \right> = \frac {a^2} {15} = \frac {1} {15} \left(\frac {\mu^* \bar{\mathcal{E}}} {kT} \right)^2,
\end{equation}
and for large electric field yields relative to thermal energies yields,
\begin{equation}\label{eq:<P2>(E)limit->infinity}
\lim_{a \rightarrow \infty} \left< P_2 \right> = 1 - \frac {3} {a} = 1 - 3 \frac {kT} {\mu^* \bar{\mathcal{E}}}.
\end{equation}

Along the same lines, the order parameter $\left< P_3 \right>$ can be calculated to give,
\begin{equation}\label{eq:<P3>(E)}
\left< P_3 \right> = - \frac {6} {a} - \frac {15} {a^3} + \left( 1 + \frac {15} {a^2} \right) \coth a .
\end{equation}
The limiting case of small electric field relative to thermal energies yields,
\begin{equation}\label{eq:<P2>(E)limit->0}
\lim_{a \rightarrow 0} \left< P_3 \right> = \frac {a^3} {105} = \frac {1} {105} \left(\frac {\mu^* \bar{\mathcal{E}}} {kT} \right)^3,
\end{equation}
and for large electric field yields relative to thermal energies yields,
\begin{equation}\label{eq:<P2>(E)limit->infinity}
\lim_{a \rightarrow \infty} \left< P_3 \right> = 1 - \frac {6} {a} = 1 - 6 \frac {kT} {\mu^* \bar{\mathcal{E}}}.
\end{equation}

\bibliography{\bibs}
\bibstyle{plain}

\end{document}